%% file: main.tex
\documentclass[runningheads]{llncs}
\usepackage{graphicx}
\usepackage{utfsym}
\usepackage{algorithm}
\usepackage{algorithmicx, algpseudocode}
\usepackage{multirow}
\newcommand{\ie}{\textit{i}.\textit{e}., }
\newcommand{\eg}{\textit{e}.\textit{g}., }
\newcommand{\who}[2]{#1 \textit{et al.} \cite{#2}}

\usepackage{chngcntr}
\usepackage[colorinlistoftodos]{todonotes}
\usepackage{hyperref}

\usepackage{color}

\begin{document}
\title{Subject-Adaptive Transfer Learning Using Resting State EEG Signals for Cross-Subject EEG Motor Imagery Classification}
\titlerunning{Subject-Adaptive Transfer Learning using RS EEG Signals}
\author{Sion An\inst{1,2} \and 
Myeongkyun Kang\inst{1} \and 
Soopil Kim\inst{1} \and 
Philip Chikontwe\inst{3} \and 
Li Shen\inst{2} \and 
Sang Hyun Park\inst{1,2} 
}


\authorrunning{An et al.}
\institute{Department of Robotics and Mechatronics Engineering, Daegu Gyeongbuk Institute of Science and Technology, Daegu, Republic of Korea \\ \email{\{sion\_an,shpark13135\}@dgist.ac.kr} \and Department of Biostatistics, Epidemiology and Informatics, University of Pennsylvania, PA, USA \and Department of Biomedical Informatics, Harvard Medical School, MA, USA }

%

\maketitle              
\begin{abstract}
Electroencephalography (EEG) motor imagery (MI) classification is a fundamental, yet challenging task due to the variation of signals between individuals \ie inter-subject variability. Previous approaches try to mitigate this using task-specific (TS) EEG signals from the target subject in training. However, recording TS EEG signals requires time and limits its applicability in various fields. In contrast, resting state (RS) EEG signals are a viable alternative due to ease of acquisition with rich subject information. In this paper, we propose a novel subject-adaptive transfer learning strategy that utilizes RS EEG signals to adapt models on unseen subject data. Specifically, we disentangle extracted features into task- and subject-dependent features and use them to calibrate RS EEG signals for obtaining task information while preserving subject characteristics. The calibrated signals are then used to adapt the model to the target subject, enabling the model to simulate processing TS EEG signals of the target subject.
The proposed method achieves state-of-the-art accuracy on three public benchmarks, demonstrating the effectiveness of our method in cross-subject EEG MI classification. Our findings highlight the potential of leveraging RS EEG signals to advance practical brain-computer interface systems. The code is available at \href{https://github.com/SionAn/MICCAI2024-ResTL}{https://github.com/SionAn/MICCAI2024-ResTL}.
\end{abstract}

\input{Docs/Introduction}
\input{Docs/Method}
\input{Docs/ExperimentsandResults}
\input{Docs/Conclusion}

\begin{credits}
\subsubsection{\ackname} This study is supported by Smart Health Care Program funded by the Korean National Police Agency (220222M01), and DGIST R\&D program of the Ministry of Science and ICT of KOREA (22-KUJoint-02), and the National IT Industry Promotion Agency(NIPA), an agency under the MSIT and with the support of the Daegu Digital Innovation Promotion Agency (DIP), the organization under the Daegu Metropolitan Government.

\subsubsection{\discintname}
The authors have no competing interests to declare that are
relevant to the content of this article.
\end{credits}

\bibliographystyle{splncs04}
\bibliography{bib}

\end{document}

%% file: Docs/Introduction.tex
\section{Introduction}
Brain-computer interface (BCI) based on electroencephalography (EEG) has emerged as a promising technology for rehabilitation~\cite{daly2008brain} and robot systems~\cite{meng2016noninvasive}, owing to their real time processing capabilities, noninvasiveness, and cost effectiveness~\cite{santhanam2006high}. In particular, EEG motor imagery (MI) signals, which represent brain activity during the imagination of body movements, are crucial in BCI research. Although prior studies~\cite{dai2020hs,song2022eeg} have achieved impressive results in intra-subject EEG MI classification, they have limited generalizability to new subjects due to the large inter-subject variability of EEG signals \ie the model may overfit to specific subjects used for training. To overcome this, several studies~\cite{9508768,yang2021novel} have proposed disentangling the extracted features into task-dependent and task-independent features. Yet, they have only partially succeeded due to insufficient information about the target subject. Alternatively, other methods use transfer learning~\cite{li2023mdtl,xie2023cross,zhang2021adaptive} and few-shot learning~\cite{an2020few,an2023dual,ng2024subject} for model adaptation by leveraging task-specific (TS) EEG signals from the target subject. However, acquiring these signals from the target subject presents practical challenges, including the extensive time costs and the difficulty in maintaining the subject's concentration, limiting their application in real-world scenarios.

Resting state (RS) EEG signals, which require a relatively more straightforward acquisition process and contain rich subject information, can provide an alternative solution for the limitations above \ie absence of the target subject information and difficulties of collecting TS EEG signals from the target subject. Despite their potential benefits, only few studies utilize RS EEG signals to decrease the domain discrepancy between multiple subjects. For example, EA~\cite{8701679} aligns TS EEG signals across all subjects using the mean of the covariance matrix computed from RS EEG signals. BCM~\cite{kwak2023subject} refines the features extracted from TS EEG signals using subject characteristics obtained from RS EEG signals. 
Although RS EEG signals enhance the model's generalizability, their models are limited in adapting to new target subject since they lack appropriate updating strategies using RS EEG signals.

To effectively address inter-subject variability of EEG data, it is essential for the model to adjust its parameters based on RS EEG signals of the target subject.
In this paper, we propose a novel subject-adaptive \textbf{Res}ting state EEG signal based \textbf{T}ransfer \textbf{L}earning \textbf{(ResTL)} for cross-subject EEG MI classification. For adapting the model to the target subject, we calibrate RS EEG signals using disentangled features to mimic TS EEG signals. Specifically, we first train a TS EEG classifier that can disentangle extracted features into task- and subject-dependent features. Afterward, we calibrate RS EEG signals to contain task-dependent features while retaining subject characteristics. Our approach is inspired by recent data-free knowledge distillation methods \cite{deepdream,yin2020dreaming} that inversely synthesize images using a pretrained classifier. We update the RS EEG signals by minimizing a combination of three different loss functions so that they exhibit task-dependent features while preserving subject-dependent features. Finally, the model is finetuned using the calibrated signals. This allows the model to be adapted to the target subject using only RS EEG signals from the target subject.

The main contributions of this work are as follows:
(i) We propose a novel transfer learning strategy that utilizes RS EEG signals for target subject model adaptation. To the best of our knowledge, this is the first attempt to leverage RS EEG signals for model adaptation in EEG MI classification. This significantly reduces the effort of collecting TS EEG signals from the target subject, requiring only RS EEG for calibration.
(ii) We calibrate RS EEG signals using the classifier to contain task-dependent features while retaining subject characteristics through incorporating feature disentanglement and inversely image synthesizing method from the pretrained classifier.
(iii) Extensive evaluation on three datasets reveals that ResTL significantly improves the accuracy of existing EEG methods for MI classification, highlighting the enhancement of the applicability in BCI applications.

%% file: Docs/Method.tex
\section{Methodology}
\begin{figure}[t!]
\centering
\includegraphics[width=\textwidth]{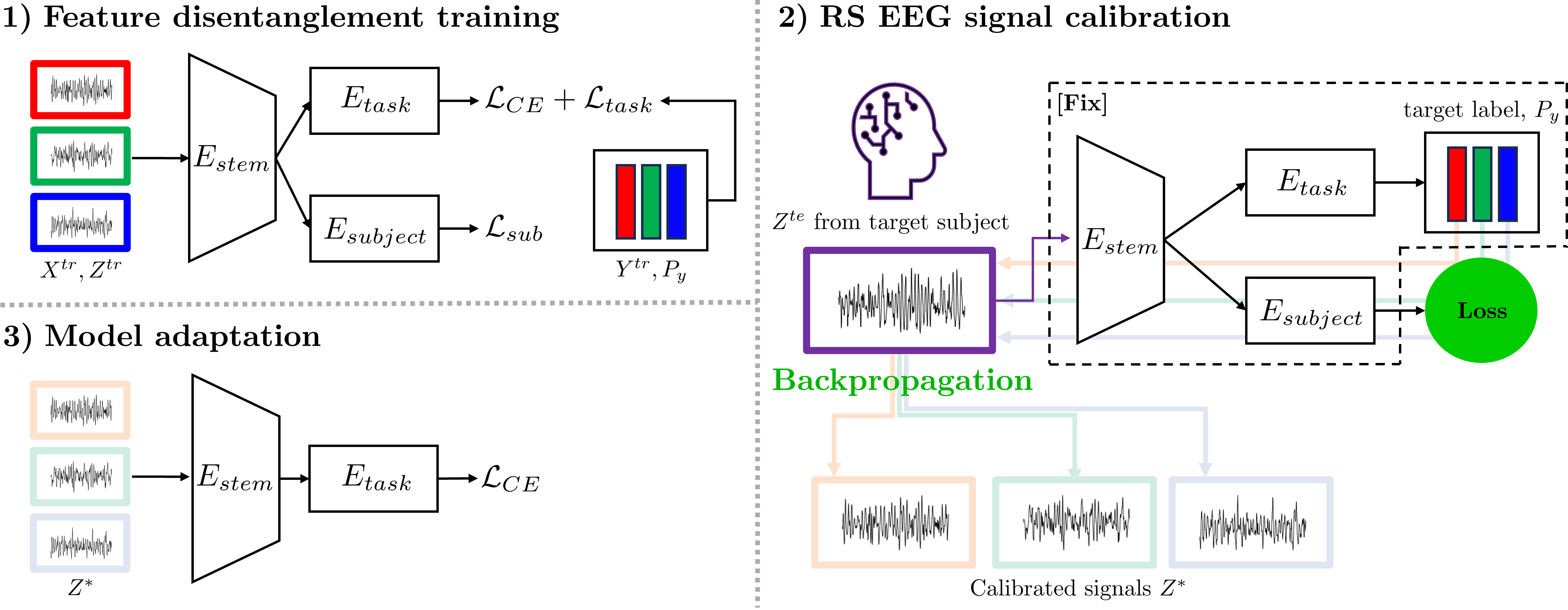}
\caption{Overview of the proposed ResTL that comprises three stages: (1) Initial classifier training using cross-entropy loss $\mathcal{L}_{CE}$, task loss $\mathcal{L}_{task}$ and subject loss $\mathcal{L}_{sub}$ for disentanglement. (2) After training, RS EEG signals (violet) are calibrated to contain task-dependent features while retaining subject-dependent features by updating the initial signals with a fixed model. Here, different colors correspond to the different class labels.
(3) Finally, the calibrated signals are used to fine-tune the pre-trained model for the target subject, minimizing $\mathcal{L}_{CE}$.}
\label{fig:framework}
\end{figure}

Our strategy (ResTL) comprises three steps as shown in Fig.~\ref{fig:framework}: (1) Feature disentanglement, (2) RS EEG signal calibration, and (3) Model adaptation. 
In the feature disentanglement training step, we train the model using TS EEG signals $X^{tr}$, corresponding labels $Y^{tr}$, and RS EEG signals $Z^{tr}$ from multiple subjects. This aims to train the classifier and disentangle extracted features into task-dependent and subject-dependent features.
In the RS EEG signal calibration step, we calibrate target subject's RS EEG signals $Z^{te}$ using the trained model. This imbues RS EEG signals with task-dependent features while preserving subject-dependent features.
In the model adaptation step, the calibrated signals $Z^*$ are used to adapt the model to the target subject. This allows the model to simulate processing TS EEG signals from the target subject by learning $Z^*$. In evaluation, the adapted model predicts the label $Y^{te}$, taking the TS EEG signals $X^{te}$ for unseen subjects.
Note that no target subject information is used during the training phase, indicating that the model can rapidly adapt to new target subjects. We present the algorithm of ResTL in Appendix.
 
\subsubsection{Feature disentanglement training} 
To obtain disentangled features, we design our model with four core components: a stem encoder $E_{stem}$, task-dependent encoder $E_{task}$, subject-dependent encoder $E_{subject}$, and classifier $C$. $E_{stem}$ extracts features from given EEG signals that are fed to both $E_{task}$ and $E_{subject}$. Our goal is to extract subject-invariant task-dependent features using $E_{task}$, and task-invariant subject-dependent features using $E_{subject}$. Additionally, the classifier $C$ predicts the final MI labels $\hat{y}$ using the task-dependent features.

To train the model, we employ two losses for $E_{task}$ and one loss for $E_{subject}$. For $E_{task}$, the cross-entropy loss $\mathcal{L}_{CE}$ is used to facilitate EEG MI classification. We further employ a task loss $\mathcal{L}_{task}$ to obtain the task-dependent features from $E_{task}$. We make prototypes $P_y$ for each class and compute the distance between the extracted features from $E_{task}$, which are then used for center loss~\cite{yang2021novel} with metric learning~\cite{schroff2015facenet} as follows: 
\begin{equation}
\label{eq:center}
\mathcal{L}_{task} = -\frac{1}{N} \sum\limits^{N}_{i=1} (d(f_i, P_{y_i}) + \sum\limits_{y_i \neq y_j} max(0,\; d(f_i, P_{y_i})-d(f_i, P_{y_j})+m)),
\end{equation}
where $i$ and $j$ are indexes of signals, $m$ is a margin for the metric learning, $d$ indicates $l_2$-norm distance, $f_i$ is a global average pooled output of $E_{task}$ and $P_{y_i}$ is a tensor representing the center of the task-dependent feature distribution corresponding to $y_i$. 

For $E_{subject}$, we employ a subject loss $\mathcal{L}_{sub}$ to obtain subject-dependent features. Specifically, we employ the triplet loss~\cite{schroff2015facenet} to minimize the distance between the anchor and positive samples while maximizing the distance between an anchor and negative samples as follows:
\begin{equation}
\label{eq:subject}
\mathcal{L}_{sub} = -\frac{1}{N} \sum\limits^{N}_{i=1} max(0, \; d(g_i, g_p)-d(g_i, g_n)+m),
\end{equation}
where $g_i$, $g_p$, and $g_n$ denote the outputs of $E_{subject}$ when fed with input samples $x_i$, $x_p$ and $x_n$, respectively.
Here, $x_p$ represents a positive EEG signal from the same subject(=$x_i$) regardless of the class, while $x_n$ represents a negative EEG signal from a different subject regardless of the class.
Additionally, we calculate $\mathcal{L}_{sub}$ using RS EEG signals by sampling positive and negative samples from $Z^{tr}$. 

Finally, the model parameters $\Theta$ in $E_{stem}$, $E_{task}$, $E_{subject}$ and $C$, are trained as follows: 
\begin{equation}
\label{eq:losstotal}
\Theta^*=\mathop{\arg \min}\limits_{\Theta} \; \mathcal{L}_{CE} + \lambda_1\, \mathcal{L}_{task} + \lambda_2\, \mathcal{L}_{sub},
\end{equation}
where $\lambda_1$ and $\lambda_2$ are hyperparameters. Meanwhile, the prototypes $P_y$ are updated by a moving average using the distance $d(f_i,P_{y_i})$ for each step as follows:
\begin{equation}
\label{eq:moving}
P_{y_i} \leftarrow P_{y_i} -\frac{\epsilon}{N} \sum\limits^{N}_{i=1} d(f_i,P_{y_i}),
\end{equation}
where $\epsilon$ is a hyperparameter. 

\subsubsection{RS EEG signals calibration.}
Generative methods~\cite{lee2023source,tian2023dual} have been proposed to augment TS EEG signals. However, they require the TS EEG signal from the target subject to reliable signal augmentation. To the best of our knowledge, generating TS EEG signals using RS EEG signals is under-explored in EEG MI tasks, and this approach may necessitate additional model training on a large dataset. Thus, we adopt the inversely image synthesis from the classifier framework for signal calibration as it does not require additional models or datasets.

To calibrate RS EEG signals and imbue them with task-dependent features while retaining subject characteristics via $E_{subject}$, we first input RS EEG signals $z \in Z^{te}$ and extract subject-dependent features $g$ that captures the unique characteristics of each subject. 
To preserve characteristics, we minimize the $l_2$-norm distance between $g$ and the extracted features from $E_{subject}$.
Moreover, we utilize trained prototypes $P_y$ to ensure that the task-dependent features extracted from the calibrated signals align with the distribution of task-dependent features via Eq.~\ref{eq:center}.
Thus, we optimize $z^*$ to contain task-dependent features while preserving subject characteristics in $z$ as follows:
\begin{equation}
\label{eq:generation}
z^* =\mathop{\arg \min}\limits_{z} \mathcal{L}_{CE} + \gamma_1\,\mathcal{L}_{task} + \gamma_2\,d(g, E_{subject}(z)),
\end{equation}
where $\gamma_1$ and $\gamma_2$ are hyperparameters. Consequently, this allows us to calibrate $K$ unique signals from a single RS signal, where $K$ represents the number of classes.

\subsubsection{Model adaptation.}
Given the calibrated signals $Z^*$, we adapt the trained model to the target subject using standard transfer learning that updates all model parameters. Thus, only $\mathcal{L}_{CE}$ is used for adaptation using calibrated signals as follows:
\begin{equation}
\label{eq:losstl}
\Theta^*=\mathop{\arg \min}\limits_{\Theta} \; \mathcal{L}_{CE}.
\end{equation}

%% file: Docs/ExperimentsandResults.tex
\section{Experimental results}
\subsubsection{Dataset.}
We evaluate ResTL on three public datasets. \textbf{BCI competition IV-2a~\cite{brunner2008bci} (BCI IV-2a)} contains raw EEG signals of nine subjects in twenty-two channels with 250Hz sampling rate for four classes \ie imaging movement of left hand, right hand, both feet and tongue. Two sessions per subject are provided, and each session consists 72 trials for each class \ie total 288 signals. 
\textbf{BCI competition IV-2b~\cite{leeb2008bci} (BCI IV-2b)} contains raw EEG signals of nine subjects in three channels with 250Hz sampling rate for two classes \ie imaging movement of left hand and right hand, and provides five sessions per subject and each session consists 200/240 trials. 
\textbf{OpenBMI~\cite{lee2019eeg}} contains raw EEG signals of fifty-four subjects in sixty-two channels with 1000Hz sampling rate for two classes \ie imaging movement of left hand right hand, and provides two sessions per subject with 200 trials each. We down-sampled EEG signals to 250Hz to match other datasets.
In all experiments, we employ 0$\sim$3s signals and 3$\sim$6s signals for RS signals and TS signals, respectively. For preprocessing, we apply a 0.5$\sim$40Hz bandpass filter with standardization. 

\subsubsection{Experimental details.}
In all experiments, we evaluate ResTL with leave-one-subject-out validation \ie one subject for testing and the remainder for training. We randomly sample 20\% of the train set as the validation set, and set $\lambda_1$, $\lambda_2$, and $\epsilon$ to 0.5, 0.05, and 1e-5, respectively. For training, we employ an Adam optimizer with 5e-4 learning rate (LR) for 100 epochs, decayed exponentially (0.99) for 10 epochs. Model selection is based on the minimum validation loss. All available RS EEG signals within each session were used to calibrate the RS EEG signals. During calibration, we set $\gamma_1$ to 1 and $\gamma_2$ to 10, using Adam optimizer with 5e-3 LR, and calibrate the RS EEG signal 300 times. For model adaptation, we fine-tune the trained model for 10 epochs.

\subsubsection{Comparison methods.}
We compare ResTL with prior works that do not use RS EEG signals~\cite{autthasan2021min2net,dai2020hs,9508768,lawhern2018eegnet,schirrmeister2017deep,song2022eeg,zhang2020motor,zhang2019convolutional}. For a fair comparison, recent work BCM~\cite{kwak2023subject} that uses RS EEG signals is included.
We use scores reported in~\cite{autthasan2021min2net,9508768,kwak2023subject}, and reproduce cases where unavailable.
Encoders EEGNet~\cite{lawhern2018eegnet} and Conformer~\cite{song2022eeg} are integrated in ResTL with the first encoder block as the stem layer $E_{stem}$, and the rest as $E_{task}$. The architecture of $E_{subject}$ is identical with $E_{task}$, except for the addition of a linear layer at the end. 

\addtolength{\tabcolsep}{5pt}
\begin{table}[t]
\caption{Classification accuracy on three datasets. Each row shows the average accuracy and standard deviation across all subjects for each comparison method. Bold font indicates the best accuracy.}
\centering
\label{tab:all}
\resizebox{0.9\columnwidth}{!}{%
\begin{tabular}{lccc}
\hline
Method                                    & BCI IV-2a          & BCI IV-2b         & OpenBMI \\ \hline
DeepConvNet~\cite{schirrmeister2017deep}  & 32.70 $\pm$ 4.30   & 68.35 $\pm$ 6.65  & 68.33 $\pm$ 15.33   \\
EEGNet~\cite{lawhern2018eegnet}           & 51.30 $\pm$ 7.36   & 67.81 $\pm$ 6.52  & 68.84 $\pm$ 14.12   \\
HS-CNN~\cite{dai2020hs}                   & 39.27 $\pm$ 6.56   & 68.52 $\pm$ 7.11  & 71.98 $\pm$ 9.15   \\
CRAM~\cite{zhang2019convolutional}        & 59.10 $\pm$ 10.74  & 66.57 $\pm$ 9.03  & 75.46 $\pm$ 10.52   \\
\who{Jeon}{9508768}                       & 58.36 $\pm$ 9.68   & 70.84 $\pm$ 6.06  & 73.32 $\pm$ 13.55   \\
MIN2Net~\cite{autthasan2021min2net}       & 53.58 $\pm$ 6.96   & 69.58 $\pm$ 7.88  & 72.03 $\pm$ 14.04   \\ 
GCRAM~\cite{zhang2020motor}               & 60.11 $\pm$ 9.96   & 69.98 $\pm$ 8.30  & 76.42 $\pm$ 9.85    \\
Conformer~\cite{song2022eeg}              & 58.23 $\pm$ 10.54  & 68.23 $\pm$ 6.65  & 77.98 $\pm$ 9.58   \\ \hline
EEGNet-BCM~\cite{kwak2023subject}         & 58.00 $\pm$ 7.61   & 69.39 $\pm$ 6.82  & 72.50 $\pm$ 14.34   \\
CRAM-BCM~\cite{kwak2023subject}           & 61.82 $\pm$ 10.68  & 70.53 $\pm$ 8.68  & 76.88 $\pm$ 10.09   \\
Conformer-BCM~\cite{kwak2023subject}      & 62.23 $\pm$ 11.73  & 71.38 $\pm$ 7.89  & 78.49 $\pm$ 10.21    \\
EEGNet-ResTL                              & 62.07 $\pm$ 7.71   & 73.13 $\pm$ 9.12  & 83.44 $\pm$ 9.06   \\
Conformer-ResTL                           & \textbf{65.09 $\pm$ 11.72} & \textbf{73.76 $\pm$ 9.41} & \textbf{86.05 $\pm$ 9.10}        \\ \hline
\end{tabular}
}
\end{table}
\addtolength{\tabcolsep}{-5pt}

\subsubsection{Main results.}
In Table~\ref{tab:all}, we present the classification results on BCI IV-2a, BCI IV-2b and OpenBMI datasets. Across all experiments, ResTL achieves the best average accuracy when employing Conformer as the encoder. Notably, the accuracy consistently improved in most cases when compared with the baselines \eg EEGNet and Conformer. This suggests that the calibrated signals are advantageous in adapting the model to the target subject. Furthermore, ResTL outperforms BCM, exceeding +2\% on three datasets. BCM utilizes RS EEG signals to correct extracted features alone, thus limiting its ability to adapt the model to the target subject. This highlights that ResTL utilizes the RS EEG signals more effectively compared to BCM. In Appendix, we report the classification accuracy for each subject on BCI IV-2a and BCI IV-2b.

\subsubsection{The distribution of the calibrated signals.}
\begin{figure}[t!]
\centering
\includegraphics[width=1.0\textwidth]{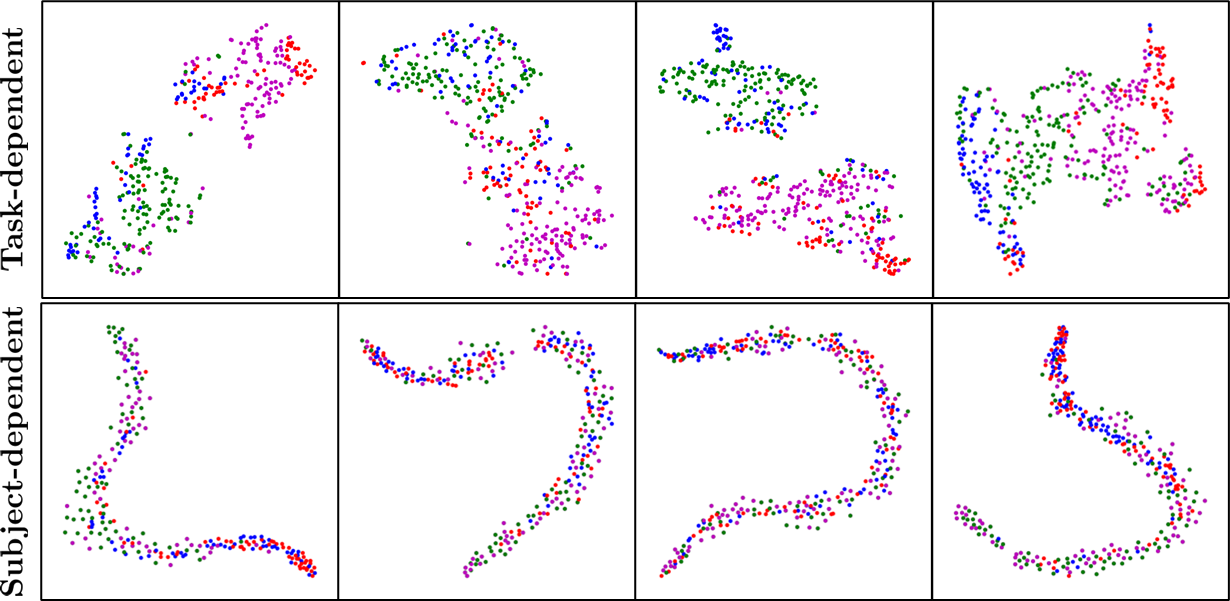}
\caption{t-SNE of calibrated signals on BCI IV-2b. (Blue: Right TS EEG signals, Red: Left TS EEG signals, Green: Calibrated signals to right, Violet: Calibrated signals to left). Calibrated signals generally have similar task-dependent features as TS EEG signals corresponding to their respective target classes \ie (Blue, Green) and (Red, Violet) are closely clustered. While all subject-dependent features are clustered in the same group, indicating similar subject-dependent features. This clearly highlights that subject characteristics in RS EEG signals are preserved during calibration.}
\label{fig:features}
\end{figure}

Fig.~\ref{fig:features} presents the t-SNE visualizations of both task-dependent and subject-dependent features extracted from the calibrated signals on the BCI IV-2b dataset. For comparison, we also visualize both features from the TS EEG signals from the target subject. The task-dependent features are clustered closely to their corresponding classes, suggesting that signals calibrated to the right exhibit similar task-dependent features as right EEG signals, and vice versa. Additionally, all subject-dependent features are clustered together regardless of their class, indicating that the calibrated signals preserve the original subject characteristics. These findings support the design of ResTL in using RS EEG signals, including performance gains. 

\subsubsection{Effect of RS EEG signals and subject-dependent features}
We compare ResTL with two inversely image synthesizing methods from the pretrained classifier \eg DeepDream~\cite{deepdream} and DeepInversion~\cite{yin2020dreaming} to assess the importance of RS EEG signals and subject-dependent features. For comparison, we synthesize TS EEG signals from both noise and RS EEG signals using each synthesis method, and then adapt the model with the synthesized signals. Table~\ref{tab:generation-ablation} presents classification results on BCI IV-2a and BCI IV-2b datasets with EEGNet and Conformer as encoders. Here, DeepDream and DeepInversion without RS EEG signals report significantly lower accuracy compared to the results with RS EEG signals. This validates our assertion that the RS EEG signals containing task characteristics are beneficial for model adaptation in EEG MI classification. Also, ResTL yields the best performance, showing the subject-dependent features can help preserving subject characteristics while calibrating RS EEG signals. 

\subsubsection{The effect of the number of the RS EEG signals.}
We present the classification accuracy corresponding to varying numbers of RS EEG signals in Table~\ref{tab:RS-ratio} on the BCI IV-2a and BCI IV-2b datasets. We observe a linear increase in performance as the number of RS EEG signals increases. This suggests that additional RS EEG signals contribute abundant information about the target subject, thereby enhancing classification performance.

\addtolength{\tabcolsep}{5pt} 
\begin{table}[t]
\centering
\caption{Comparison with TS EEG signal synthesis methods when (EEGNet/Conformer) are employed. RS and SF indicate whether to use the RS signals and the subject-dependent features, respectively.}
\label{tab:generation-ablation}
\resizebox{0.75\columnwidth}{!}{%
\begin{tabular}{ccccc}
\hline
Method        & RS & SF & BCI IV-2a & BCI IV-2b \\ \hline
DeepDream     &              &              & 57.00/61.40 & 67.51/69.32    \\
DeepDream     & \usym{1F5F8} &              & 59.86/62.13 & 72.10/71.16    \\
DeepInversion &              &              & 49.12/50.64 & 65.15/70.77    \\
DeepInversion & \usym{1F5F8} &              & 58.79/62.55 & 71.77/71.17    \\
ResTL          & \usym{1F5F8} & \usym{1F5F8} & \textbf{62.07}/\textbf{65.09} & \textbf{73.13}/\textbf{73.76}    \\ \hline
\end{tabular}%
}
\end{table}
\addtolength{\tabcolsep}{-5pt} 

\addtolength{\tabcolsep}{3pt} 
\begin{table}[t]
\centering
\caption{Classification accuracy of ResTL with (EEGNet/Conformer) corresponding to varying numbers of RS signals. The numbers in each column of the first row represent the percentile of the total number of RS EEG signals used.}
\resizebox{\columnwidth}{!}{%
\begin{tabular}{cccccc}
\hline
Dataset    & 20\%  & 40\%  & 60\%  & 80\%  & 100\% \\ \hline
BCI IV-2a & 60.07/63.33 & 60.60/63.89 & 61.19/64.59 & 61.46/64.73 & \textbf{62.07}/\textbf{65.09} \\
BCI IV-2b & 72.58/72.88 & 72.66/73.28 & 72.90/73.35 & 73.02/73.58 & \textbf{73.13}/\textbf{73.76} \\ \hline
\end{tabular}
}
\label{tab:RS-ratio}
\end{table}
\addtolength{\tabcolsep}{-3pt}

%% file: Docs/Conclusion.tex
\section{Conclusion}
We present a novel subject-adaptive transfer learning strategy for EEG MI classification, leveraging the calibration of RS EEG signals. Our approach comprises feature disentanglement and subject-dependent feature constraint during calibration. Subsequently, we fine-tune the model using the calibrated signals to adapt it to the target subject.
Extensive experiments demonstrate the effectiveness of our approach in improving the applicability of EEG MI classification across subjects, particularly in scenarios where TS EEG signals of the target subject are unavailable. 
In future studies, we aim to explore the integration of meta-learning strategies, especially in scenarios where only a few target subject samples are available. 